\documentclass[%
 aip,
 amsmath,amssymb,
 reprint,%
]{revtex4-1}

\usepackage{graphicx}% Include figure files
\usepackage{dcolumn}% Align table columns on decimal point
\usepackage{bm}% bold math
\usepackage[utf8]{inputenc}
\usepackage[T1]{fontenc}
\usepackage{mathptmx}

\begin{document}

\preprint{AIP/123-QED}

\title[]{Solitary structure formation and self-guiding of electromagnetic beam in highly degenerate electron plasma}
% Force line breaks with \\

%\author{V.I. Berezhiani,$^{1,2}$,  SOSO,$^{2}$ and Z. Osmanov}

\author{V.I. Berezhiani}
%\altaffiliation[Also at ]{Andronikashvili Institute of Physics (TSU), Tbilisi \ 0177, Georgia.}%Lines break automatically or can be forced with \\
\affiliation{ 
School of Physics, Free university of Tbilisi, Tbilisi 0159, Georgia.%\\This line break forced with \textbackslash\textbackslash
}%
\affiliation{%
Andronikashvili Institute of Physics (TSU), Tbilisi \ 0177, Georgia.%\\This line break forced% with \\
}%

\author{Z.N. Osmanov}
 \homepage{Corresponding author}
  \email{z.osmanov@freeuni.edu.ge}

% \homepage{http://www.Second.institution.edu/~Charlie.Author.}
\affiliation{ 
School of Physics, Free university of Tbilisi, Tbilisi 0159, Georgia.%\\This line break forced with \textbackslash\textbackslash
}%
\affiliation{%
E. Kharadze Georgian National Astrophysical Observatory, Abastumani 0301, Georgia.%\\This line break forced% with \\
}%

\author{S.M. Mahajan}
% \homepage{http://www.Second.institution.edu/~Charlie.Author.}
\affiliation{ 
Institute for Fusion Studies, The University of Texas at
Austin, Austin, TX 78712, USA.%\\This line break forced with \textbackslash\textbackslash
}%

\author{S. V. Mikeladze}%
%\email{Second.Author@institution.edu.}
\affiliation{%
Andronikashvili Institute of Physics (TSU), Tbilisi \ 0177, Georgia.%\\This line break forced% with \\
}%

\date{\today}% It is always \today, today,
             %  but any date may be explicitly specified

\begin{abstract}
In the present paper we consider the nonlinear interaction of high frequency intense electromagnetic (EM) beam with degenerate electron plasmas. In a slowly varying envelop approximation the beam dynamics is described by the couple of nonlinear equations for the vector and scalar potentials. Numerical simulations demonstrate that for an arbitrary level of degeneracy the plasma supports existence of axially symmetric 2D solitons which are stable against small perturbations. The solitons exist if the power trapped in the structures, being the growing function of soliton amplitude, is above a certain critical value but below the value determining by electron cavitation. The robustness of obtained soliton solutions was verified by simulating the dynamics of initial Gaussian beams with parameters close to the solitonic ones. After few diffraction lengths the beam attains the profile close to the profile of the ground state soliton and propagates for a long distance without detectable distortion. The simulations have been performed for the input Gaussian beams with parameters far from ground state solutions. It is shown that the beam parameters are oscillating near the parameters of the ground soliton solution and thus the formation of oscillating waveguide structures takes place.
\end{abstract}

\maketitle

%%%%%%%%%%%%%%%
\section{Introduction}
%%%%%%%%%%%%%%%

Multiwavelength observations on highly luminous compact astrophysical objects confirm that the detected  spectral pattern covers almost the whole interval of electromagnetic radiation \cite{Begelman}. It is worth noting that interiors of these objects are highly dense and therefore, to study physical processes in such environments one should take into account the Fermi-Dirac statistics if the corresponding Fermi energy prevails over the binding energy of electrons \cite{shapiro}. This in turn, leads to an efficient ionisation process of atoms resulting in extremely high electron gas densities $10^{26-34}$ cm$^{-3}$ when the average spacing between the particles is significantly smaller than the corresponding thermal de Broglie length-scale \cite{Landau}.

In such a dens ambient if the number density of electrons is smaller than the critical number density, $n_{c}=m_{e}^{3}c^{3}/3\pi
^{2}\hbar^{3}=5.9\times10^{29}cm^{-3}$, the rest mass energy will be small compared to the corresponding Fermi energy $\epsilon_{F}=m_{e}c^{2}\left(  \gamma_{F}-1\right)$ indicating relativistic nature of the electron gas. Here, $m_e$ denotes the electron's mass, $c$ is the speed of light, $\hbar$ denotes the Planck's constant and $\gamma_F = \sqrt{1+p_{F}^{2}/m_{e}^{2}c^{2}}$ is the so-called Fermi Lorentz factor, where $p_{F}=m_{e}c\left(
n/n_{c}\right)  ^{1/3}$ represents the Fermi momentum \cite{Haas} and $n$ is the electron's number (proper) density.

The study of non-linear interaction of high frequency electromagnetic (EM) radiation might be significant in a certain class of astrophysical objects. Generally speaking, according to observations it is assumed that $X$-ray emission from the white dwarfs (WD) is a direct result of an accretion matter, falling onto the WD's surface and via the Bremsstrahlung the high frequency EM radiation is generated \cite{wd}. This emission penetrates the interiors of the star, which predominantly is composed of a highly degenerate electron gas and non-linearly interacts with it. The study of interaction of the accretion driven $X$-ray emission with the internal structures of neutron stars (NS), where apart from neutrons there is almost $1\%$ of degenerate electrons \cite{shapiro} - seems to be very promising. The same scenario might occur in gamma ray bursters (GRB) where high energy EM radiation is a result of the supernova explosion, when massive stars finally collapse to NSs \cite{carroll}.

The laboratory study of matter under shock compression should be important in this context. In particular, to simulate the physical conditions in the interiors of giant planets, brown dwarfs and low mass stars, high intensity lasers are used to achieve high densities by means of interaction of EM radiation with matter \cite{azechi,eliezer,plasmons}. The high-power laser systems are capable of producing extremely compressed state of matter. For instance, in, \cite{azechi} it has been demonstrated that due to the laser-driven impositions of spherical polymer target its density grows up thousand times exceeding the solid density. Created in this condition plasma could be in a strongly degenerate state \cite{eliezer}. Such plasmas can be investigated by modern free electron X-ray laser facilities. Due to the modern achievements it possible to increase the laser intensities up to $10^{20}$ W cm$^{-2}$ at $9.9$ keV \cite{mimura} and improvement is still ongoing, making it possible to study the interaction of $X$-rays with degenerate a matter.

A series of works have been dedicated to the study of nonlinear interactions of high frequency radiation and plasma waves in a degenerate media \cite{Korakis,SMB,Guga,Nana1,Nana}. In particular, in \cite{SMB,Guga,Nana1,Nana} the authors have studied solitary solutions in a fully degenerate relativistic as well as non-relativistic multi-species plasmas and was shown that such structures exist regardless of the  degeneracy level. The stimulated scattering has been examined in \cite{Gio-Raman,misra}, where the instability occurs for weakly as well as a strongly degenerate electron plasma.

Recently the linear regime of the filamentation instability of the electromagnetic beam was investigated in a highly relativistic electron plasma  \cite{GBO}. As it was found, the instability arises in a weakly as well as strongly relativistic plasma. The growth rate of the instability was estimated and analysed for an arbitrary level of degeneracy and the critical power trapped in a single filament has been estimated.

In the present paper we consider the non-linear formation of $2D$ solitary structures in unmagnetised electron plasmas and study the stability problem for a variety of parameters. The similar problem but for the electron-positron plasmas was addressed in \cite{Nana1}. Since in the electron-positron plasmas charge separation does not occur, the governing set of equations reduce to a single equation with saturated non-linearity for a vector potential in parabolic approximation describing dynamics of the EM beam. In contrast, in electron plasmas the governing equations reduce to a couple of non-linear equations derived in \cite{GBO}. 

The paper is organized in the following way: in Sec. 2 we introduce the governing equations, numerically solve them and obtain corresponding results and in Sec. 3 we briefly outline them.

%%%%%%%%%%%%%%%
\section{Main Consideration}
%%%%%%%%%%%%%%%

As it has been shown in \cite{GBO} the dynamics of relativistic EM beams
propagating in a degenerate electron plasma can be described by the set of
nonlinear equations which in a dimensionless form reads:

\begin{equation}
2i\frac{\partial A}{\partial z}+\nabla _{\perp }^{2}A+\left( 1-\frac{N}{%
1+\Psi }\right) A=0  \label{A1}
\end{equation}

\begin{equation}
\nabla _{\perp }^{2}\Psi +1-N=0  \label{psi1}
\end{equation}

\begin{equation}
N=\frac{\left( 1+\Psi \right) \left[ \left( 1+\Psi \right) ^{2}-\left(
1+\left\vert A\right\vert ^{2}-d\right) \right] ^{3/2}}{d^{3/2}\left[ \left(
1+\Psi \right) ^{2}-\left\vert A\right\vert ^{2}\right] ^{1/2}}  \label{N1}
\end{equation}

Here $A$ is the slowly varying amplitude of the circularly polarized vector
potential $e\mathbf{A/}\left( m_{e}c^{2}\Gamma _{0}\right) \mathbf{=}\left( 
\widehat{\mathbf{x}}+i\widehat{\mathbf{y}}\right) A\exp (-i\omega
_{0}t-k_{0}z)+c.c.$, where $\widehat{\mathbf{x}}$ and $\widehat{\mathbf{y}}$
are the unite vectors directed across the EM beam propagation direction $z$ . The
field frequency $\omega _{0}$ and the wave vector $k_{0}$ satisfy the dispersion
relation $\omega _{0}^{2}=k_{0}^{2}c^{2}+\omega _{e}^{2}/\Gamma _{0}$ , $%
\omega _{e}=\left( 4\pi e^{2}n_{0}/m_{e}\right) ^{1/2}$ - is the plasma
frequency, $n_{0}$ is the equilibrium density of electrons and $\Gamma
_{0}=\left( 1+R_{0}^{2}\right) ^{1/2}$ is the generalized relativistic
factor, $R_{0}=\left( n_{0}/n_{c}\right) ^{1/3}$. \ $\Psi =e\varphi
/\left( m_{e}c^{2}\Gamma _{0}\right) $ where $\varphi $ is the charge
separation scalar field which is created due to the action of the high-frequency
pressure on electrons. $N=N/n_{0}$ is the normalized electron density while
the dimensionless coordinates read as $z=\left( \omega _{e}^{2}/c\omega
_{0}\Gamma _{0}\right) z$, $\mathbf{r}_{\perp }=\left( \omega _{e}/c\sqrt{%
\Gamma _{0}}\right) \mathbf{r}_{\perp }$. In Eq. (\ref{N1}) $d=R_{0}^{2}/\left(
1+R_{0}^{2}\right) $ is a measure of the level of degeneracy. For a weakly degenerate
case $\left( R_{0}<<1\right) $ $d\simeq R_{0}^{2}$ while for relativistic
degeneracy $\left( R_{0}>>1\right) $ $d\rightarrow 1$.

Deriving the system of Eqs. (\ref{A1}-\ref{N1}) it is assumed that plasma is highly
transparent $\omega _{e}/\omega _{0}<<1$ and $\lambda <<L_{\perp
}<<L_{\parallel }$ where $\lambda \approx 2\pi c/\omega _{0}$ is the
wavelength of EM radiation, $L_{\perp }$ and $L_{\parallel }$ are the
characteristic longitudinal and transverse spatial dimensions of the EM
beam. This system describes the dynamics of the strong amplitude narrow EM beams
in plasmas with an arbitrary (but physically justified) strength of degeneracy.

Bases on the system of Eqs. (\ref{A1}-\ref{N1}) we investigate numerically the
possibility of self-trapping of intense electromagnetic beam to demonstrate
the formation of stable 2D solitonic structures in such plasmas. The only
external parameter in the system (\ref{A1}-\ref{N1}) is the level of degeneracy $d$.
At this end we would like to emphasize that for the degenerate plasma the
average energy of the charge particle interaction should be less than the Fermi
energy. This condition implies that plasma electron density should be $%
n_{0}\geq e^{6}m_{e}^{3}/\hbar ^{6}=\allowbreak 6.\,7\times 10^{24}cm^{-3}$
\ and consequently $R_{0}>>0.02$ , note that for $n_{0}=n_{c}=5.9\times
10^{29}cm^{-3},$ $R_{0}=1$ while $n_{0}=10^{34}cm^{-3}$ corresponded to the
ultra-relativistic degeneracy with $R_{0}\approx 26$. \ For the electron
densities $10^{26}cm^{-3}-10^{34}cm^{-3}$ the level of the degeneracy measure $d$
in Eq.(\ref{N1}) varies in the range $0.003<d<0.999$. Since $d$ is rather small for
the nonrelativistic densities, we can safely assume that $d\rightarrow 0$ in
Eqs.(\ref{A1}-\ref{N1}) and then obtain the following relations:

\begin{equation}
\Psi =\sqrt{1+\left\vert A\right\vert ^{2}}-1  \label{psi2}
\end{equation}

\begin{equation}
N=1+\nabla _{\perp }^{2}\sqrt{1+\left\vert A\right\vert ^{2}}  \label{N2}
\end{equation}%
reducing the system of Eqs.(\ref{A1}-\ref{N1}) to the single equation for $A$

\begin{equation}
2i\frac{\partial A}{\partial z}+\nabla _{\perp }^{2}A+\left( 1-\frac{1}{%
\sqrt{1+\left\vert A\right\vert ^{2}}}-\frac{\nabla _{\perp }^{2}\sqrt{%
1+\left\vert A\right\vert ^{2}}}{\sqrt{1+\left\vert A\right\vert ^{2}}}%
\right) A=0  \label{A2}
\end{equation}%
An equation similar to Eq. (\ref{A2}) has been derived in past for cold classical plasmas
and widely exploited for the problem of relativistic self-focusing of the
laser beams \cite{sun}. It was shown that Eq.(\ref{A2}) supports existence of radially
symmetric 2D solitonic structures, which are stable against small
perturbations. Thus, even if the density of the weakly degenerate plasma is few
order magnitude larger than laser plasma density the results of \cite{sun} are
fully applicable in our case. However, Eq.(\ref{N2}) indicates that at high
intensity or for radially strongly localized beams the electron density $N$\
may become zero or even negative. Zero density or the complete expulsion of
electrons (electron cavitation) is possible in the region of the strong
field because the charge separation electric field cannot oppose the
ponderomotive force of EM field. In this paper, for the soliton solutions described below we
consider the domain of parameters when the condition $N>0$\ holds, i.e., the
electron cavitation does not take place \cite{BMYP}.

For the finite values of the degeneracy parameter $d$ one has to solve a more
complex system of equations (\ref{A1}-\ref{N1}) numerically. To establish the possibility
of existence of the 2D solitonic -ground state structure we look for solutions
which are radially symmetric: $A=A(r)\exp \left( ik/2\right) $, $\Psi
=B\left( r\right) ,$ here $k$ is a propagation constant, and fields are
assumed to be localized, i.e. for $r=\sqrt{x^{2}+y^{2}}\rightarrow \infty $, $%
A\rightarrow 0,B\rightarrow 0$. For the ultrarelativistic degeneracy level $%
d\rightarrow 1$ the governing set of a couple of nonlinear ordinary
differential equation is:

\begin{equation}
\frac{1}{r}\frac{d}{dr}\left( r\frac{dA}{dr}\right) -kA+A\left( A^{2}-\left(
1+B\right) ^{2}\right) =0  \label{A}
\end{equation}

\begin{equation}
\frac{1}{r}\frac{d}{dr}\left( r\frac{dB}{dr}\right) +1+\left( 1+B\right)
\left( A^{2}-\left( 1+B\right) ^{2}\right) =0  \label{B}
\end{equation}

\begin{figure}
 \centering
  \resizebox{8cm}{!}{\includegraphics[angle=0]{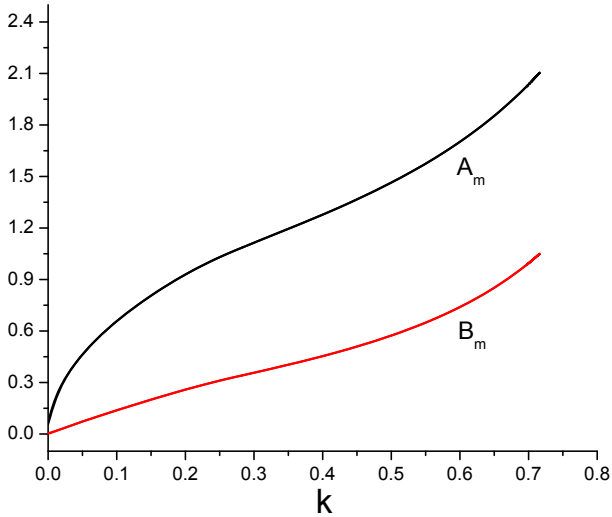}}
  \caption{The field amplitudes $A_{m}$ and $B_{m}$ versus the propagation
constant $k$ for the ultrarelativistic degenerate plasma ( $d=1$).} \label{fig1}
\end{figure}

\begin{figure}
 \centering
  \resizebox{8cm}{!}{\includegraphics[angle=0]{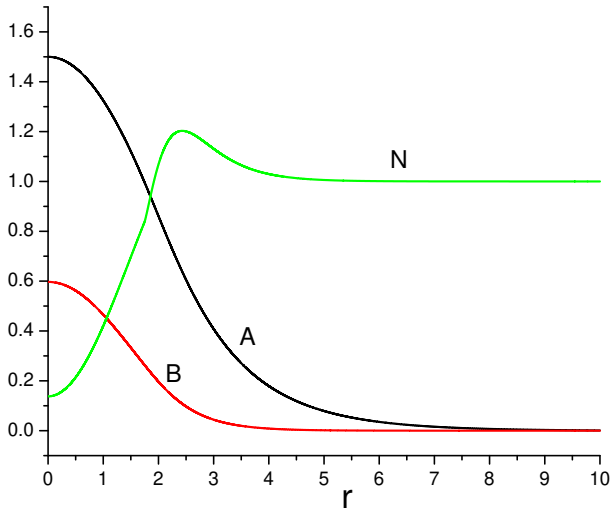}}
  \caption{The profiles of the field potentials and the electron plasma density
for $d=1$ and $k\approx 0.52$.} \label{fig2}
\end{figure}

Applying the shooting code we solved the eigenvalue problem of Eqs. (\ref{A}-\ref{B}) for
the fundamental, noddles localized solutions when the fields attain their
maxima $A_{m}$ , $B_{m}$ at $r=0$ and smoothly decaying for $r\rightarrow
\infty $. Such solutions exist if the propagation constant that is
the eigenvalue of the problem is in the range $0<k<k_{c}\simeq 0.72$. In Fig.1
dependence of $A_{m}$ $\ $and $B_{m}$ on $k$ is displayed. One can see that
amplitudes of both fields are growing functions of the propagation constant.
For $k\lesssim 0.3$ EM field amplitude is $A_{m}<1$ while for $k\gtrsim 0.3$
the amplitude becomes ultrarelativistic $A_{m}\gtrsim 1$. Profiles of the
solutions are similar for different $k$ and for $k\approx 0.52$ as can be seen
in Fig.2. Amplitudes of the fields are $A_{m}=1.5,$ $B_{m}=0.58$ and as one
could expect the action of ponderomotive force leads to electron density
reduction at the center of the structure $N\left( 0\right) =0.15$. For $k$
approaching a critical value $k_{c}$ \ the electron density $N\left(
0\right) $ tends to zero, i.e. the electron cavitation takes place. In the regime 
corresponding to cavitation the amplitudes of the fields are $%
A_{m}=2.1,$ $B_{m}=1.05$. For larger values of $k(>k_{c})$ the electron
density becomes negative and any model of plasma which is based on the fluid
description is not applicable \cite{fluid}. Thus, relativistic
degenerate plasma supports existence of the localized 2D solitary beam with EM
field amplitude bounded from above $A_{m}<2.1$.

\begin{figure}
 \centering
  \resizebox{8cm}{!}{\includegraphics[angle=0]{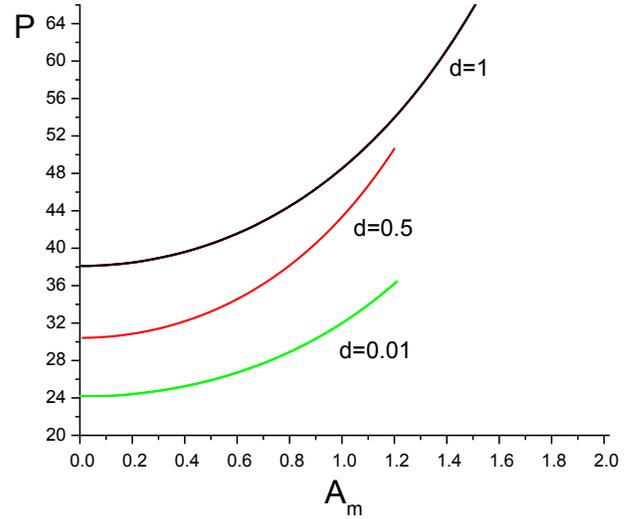}}
  \caption{The power $P$ vs $A_{m}$ for different $d$: $d<<1$, $d=0.5$ and $d=1$.} \label{fig3}
\end{figure}

We have performed the similar simulations for different values of the degeneracy
parameter in the range $1>d>0.003$ and  established that field amplitudes
are growing functions of $k$. At the same time the EM field and the plasma
density profiles qualitatively resemble the one shown in Fig.2 and for
brevity we do not present corresponding plots here. However, we would like
to emphasize that with decreasing $d$ the critical value of $k_{c}$ as well
as the corresponding field amplitude bound is also decreasing. For instance for $%
d=0.5$ the EM field amplitude is bounded from above $A_{m}<1.8$. For $d=0.003
$ we applied Eq. (\ref{N1}) and found that $A_{m}<1.5$.

The EM beam power $\int d\mathbf{r}_{\perp }\left\vert A\right\vert ^{2}$ is
an integral of motion of the system of Eqs. (\ref{A1}-\ref{N1}). Let us define the dimensionless
power of the EM beam that is trapped in self-guided steady 2D solitonic
structures as

\begin{equation}
P=2\pi \int_{0}^{\infty }drrA^{2}  \label{P}
\end{equation}%
As it follows from numerical simulations $P$ is a growing function of the
propagation constant $k$\textbf{. }Since $k$ is also a growing function of $%
A_{m}$ (see above) in Fig. 3 we exhibit the dependence of trapped power on $%
A_{m} $ for different values of the degeneracy parameter $d$. We can see
that for weak as well as for strongly degenerate cases the trapped power in
2D solitons are above a certain critical value $P_{0}$ that is reached for
small amplitudes $A_{m}<<1$ while an upper bound of the power $P_{c}$ is
related to the electron cavitation that is taking place for larger $A_{m}$.
In particular, for $d<<1,P_{0}\simeq 23.9$ and $P_{c}\simeq 27.9$ while for $%
d\simeq 1$ both bounds of the powers are increasing $P_{0}\simeq 34.2$, $%
P_{c}\simeq 57.9$. Here we remark that for $d<<1$ the critical power $%
P_{0}\simeq 23.9$ coincides with the one obtained in \cite{sun} and in
dimensions reads as $P_{0}\simeq 16.2\left( \omega /\omega _{e}\right)
^{2}GW $.

\begin{figure}
 \centering
  \resizebox{8cm}{!}{\includegraphics[angle=0]{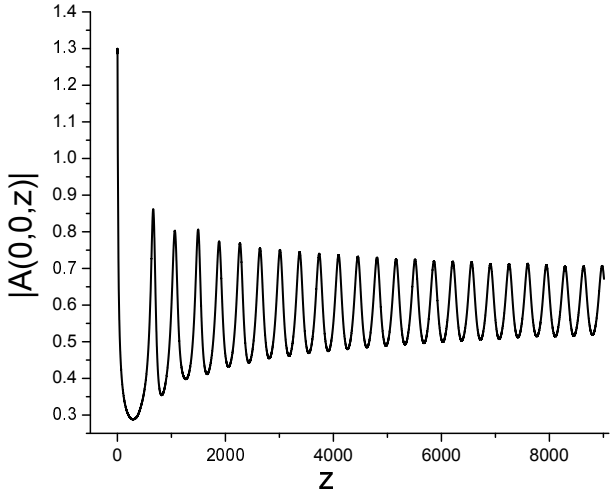}}
  \caption{The field intensity $\left\vert A\left( 0,0,z\right) \right\vert $ vs
the propagation distance $z$ for the Gaussian beam with the input amplitude $A_{0}=1.3$
and power $P=48$ for  $d=1$.} \label{fig4}
\end{figure}

Main issue we have to address now is the stability of the obtained solitary
solutions. Since for weak degeneracy $\left( d<<1\right) $ Eq. (\ref{A2}) is
applicable we can conclude that the solitary solutions are stable in this
limit. To investigate the stability of the solutions for the finite values
of $d$ we conducted the numerical simulations for the full system of Eqs.
(\ref{A1}-\ref{N1}) based on the finite difference scheme with soft-boundary conditions. The
initial conditions (at $z=0$) for our simulations were numerically obtained
solutions (similar to one exhibited in Fig.2) for different $d$ and for the
power in the range $P_{0}<P<P_{c}$ (i.e. the field amplitude $A_{m}$ is
varying from nonrelativistic to ultrarelativistic values). Because a finite
difference scheme approximates partial differential equations with only
limited accuracy, using the numerically obtained 2D solution as an initial
condition implies an inherent initial error (perturbation). The level of the
perturbation depends on the number of points on the mesh. Extensive
simulations carried out up to the propagation distance $z=15\;z_{dif}$ \ where 
$z_{dif}=2\pi L_{\perp }^{2}$ - is the linear diffraction length of EM beam.
\ Changing the number of points did not affect the shape of solitary
structures during an entire time of simulation. Thus, we conclude that the
solitary structures are stable against small perturbations.

\begin{figure}
 \centering
  \resizebox{8cm}{!}{\includegraphics[angle=0]{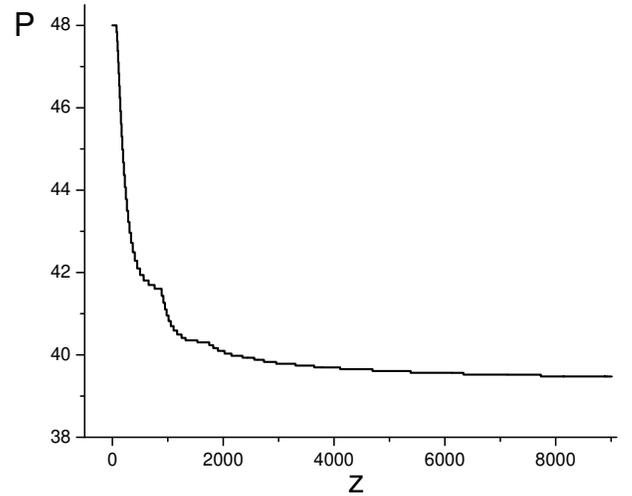}}
  \caption{Corresponding (to Fig.4) power decrease due to radiative losses} \label{fig5}
\end{figure}

The next question to investigate is the nonlinear stability when the initial
shape and strength of the EM field or its fluctuations are far from the
shape and strength of the soliton solution. For this purpose an input beam
is assumed to be Gaussian $A(x,y,0)=A_{0}\exp \left[ -\left(
x^{2}+y^{2}\right) /2l^{2}\right] $ where $A_{0}$ and $l=A_{0}^{-1}\sqrt{%
P/\pi }$ respectively are the amplitude and the characteristic width of the
beam. In the series of simulations conducted for different $d$\ the beam
power is in the range $P_{0}<P<P_{c}$ while the amplitude of the beam is taken
to be nearly equal to the amplitude of steady (equilibrium) solutions $%
A_{0}\approx A_{m}$ . In all considered cases parameters of the Gaussian
beam undergo small but damped oscillations around parameters of solitary
solutions while the corresponding power slightly decreases due to radiative
losses. After a few diffraction lengths the beam attains the profile close to
the profile of the ground state soliton and propagates for a long distance
without detectable distortion.

\begin{figure}
 \centering
  \resizebox{8cm}{!}{\includegraphics[angle=0]{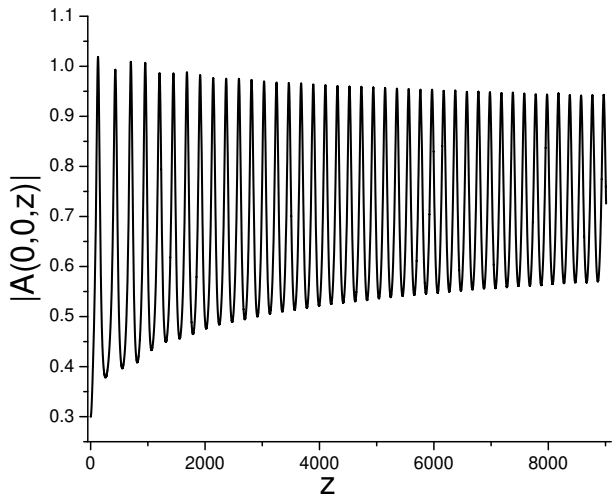}}
  \caption{The field intensity $\left\vert A\left( 0,0,z\right) \right\vert $ vs
the propagation distance $z$ for the Gaussian beam with the input amplitude $A_{0}=0.3$
and power $P=48$ for $d=1$. } \label{fig6}
\end{figure}

Now question is what happens if the power and amplitude of the input
Gaussian beam are far from parameters of ground state solitary solutions.
Our simulations show that if the power of the beam is less than the critical one 
$P_{0}$ the beam diffracts. However if input power of the beam (at $z=0)$ is
in the range $P_{0}<P<P_{c}$ the beam can be trapped in the self-guiding regime
of propagation. In Fig. 4 we plot the evolution of field intensity $%
\left\vert A\left( 0,0,z\right) \right\vert $ for the case of relativistic degeneracy $d=1$. The input power of the beam is taken to be $P=48$ and the
amplitude is assumed to be relativistic $A_{0}=1.3$. Simulations have been performed up to $z=160z_{dif}$. The beam undergoes structural changes with
initial defocusing which follows by the formation of the oscillatory waveguide
structure. The power in the region of the beam localization remarkably decreases
due to the generated radiation spectra (see Fig. 5). In few oscillations the
power drops from $P=48$ to $P\approx 40$ and subsequently only a weak decrease
of power can be seen having a tendency to reach the value $P=39$. We can
conclude that the beam parameters are osclilating near the parameters of
the ground soliton solution with $P\approx 39-40$. Similar simulations are
carried out for the same power $P=48$ but different amplitudes of the input beam
provided that $A_{0}<1.7$. For larger amplitude beams the electron
density becomes negative and consequently our model equations become invalid. In Fig. 6 we
plot the evolution of the field intensity $\left\vert A\left( 0,0,z\right)
\right\vert $ for the weakly relativistic input beam with $A_{0}=0.3$\textbf{. }%
The pattern of the beam dynamics is the same as for relativistic amplitudes,
but in contrast, the profile reshaping begins with beam focusing. A typical example of the spatial dynamics of the beam as well as the plasma density is plotted in Fig. 7 for different values of the propagation distance. As one can clearly see on this figure the intensity as well as the density profiles undergo an oscillatory character of dynamics.

\begin{figure}
 \centering
  \resizebox{8cm}{!}{\includegraphics[angle=0]{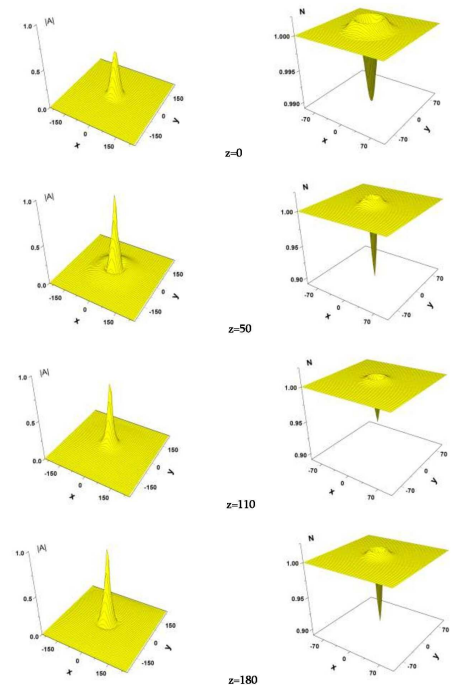}}
  \caption{ Snapshots of EM field intensity $|A|$ (left column) and plasma density
N (right column) for diiffenet propagation distances $z = 0; 50; 110; 180$ for
$P = 48$ and $A_0 = 0.6$.} \label{fig7}
\end{figure}

\begin{figure}
 \centering
  \resizebox{8cm}{!}{\includegraphics[angle=0]{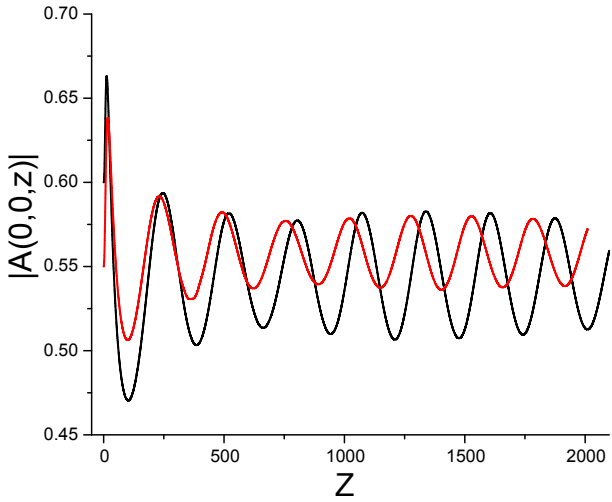}}
  \caption{The field intensity $\left\vert A\left( 0,0,z\right) \right\vert $ vs
the propagation distance $z$ for two Gaussian beams with input amplitudes $%
A_{0}=0.55$ and $A_{0}=0.6$. The input powers of both beams are the same  $P=40$
while the degeneracy parameter is $d=0.5$.} \label{fig8}
\end{figure}

The numerical simulations was conducted not only for the case of the ultrarelativistic
degeneracy ($d=1$) but also for a variety of values of the degeneracy parameter.
In Fig. 8 we present the results of simulations for $d=0.5$ (i.e. $%
n_{0}=5.9\times 10^{29}cm^{-3}$) where dynamics of the peak intensity is plotted
versus the propagation distance of two beams with slightly different input
amplitudes $A_{0}=0.55,0.6$ and with the same power $P=40$. One can see that both beams are trapped in self-guided regime of propagation with parameters oscillating around the parameters of the ground solitonic solution.

%%%%%%%%%%%%%%%
\section{Conclusion}
%%%%%%%%%%%%%%%

We have considered the possibility of the intense EM beam
localization in the degenerate transparent electron plasma. Dynamics of the
beams which is described by the system of equations (\ref{A1}-\ref{N1}), is studied
numerically for a different level of the plasma degeneracy. It is shown that for
the arbitrary level of degeneracy plasma supports existence of axially
symmetric 2D ground state solitonic structures which are stable against
small perturbations. The EM field power that is trapped in such solutions
being above the critical one ($P_{0}$) is a growing function of the soliton
amplitude and terminates at a certain amplitude (power $P_{c}$). Further growth
of the amplitude leads to electron cavitation.

The robustness of the ground state solutions have been verified introducing in
the system the Gaussian input beams with a given power and amplitude close to
the amplitude for the solitonic solutions. The beam profile undergoes small but
damped oscillation around the ground state solitonic solutions. If the amplitude
of the beam is remarkably different from the amplitude of the soliton the beam
is trapped in the self-guiding regime of propagation. In a few diffraction lengths
the beam power reduces at a certain level (but remains in the range $%
P_{0}<P<P_{c}$) due to the radiation losses. Subsequently the radiation losses
reduce considerably and the EM beam parameters oscillate around the stable
ground state solution for a long distance of propagation exceeding the diffraction length
by two orders of magnitude. Such a behavior of the beam is observed
for all level of degeneracy and for the entire range of the allowed power $%
\left( P_{0}<P<P_{c}\right) $. 

We would like to emphasize that a high power beam could undergo the
filamentation instability that leads to the break up of the field into a number of
beamlets carrying power $\sim P_{0}$. As one can see in Fig.1 the above considered models allowed power $P<2P_{0}$ for all level of degeneracy.
Consequently we did not observe the filamentation of the field structure.
For high power EM fields the electron cavitation may take place which leads
to the appearance of an area with negative electron densities. In a cold plasma case
such a failure of the hydrodynamic plasma model is generally corrected by
putting $N=0$ in the entire spatial region where $N<0$ \cite{sun,borisov}. Workability of
such an ansatz were confirmed by PIC simulations \cite{naseri} also. Similar approach can
be applied for the degenerate plasma as well however it is beyond of the
intended scope of the current paper.

The physical regimes considered in this paper might take place in a wide class of astrophysical objects: white dwarfs, neutron stars and GRBs, where the order of magnitude of the number density of electrons is respectively $10^{33}$cm$^{-3}$, $10^{36}$cm$^{-3}$ and $10^{30-37}$ cm$^{-3}$ (see for details \cite{shapiro,grb}), which indicates high degeneracy of the electron plasma. On the other hand, it is observationally evident that all mentioned objects are characterised by high intensities of EM radiation ($X$-rays and/or $\gamma$-rays), therefore, the physical conditions considered in this paper might be significant in the mentioned objects. To study particular realisations of the conditions one has to examine individual objects, which is beyond the intended scope of the present paper. 

\begin{acknowledgments}
The work was supported by the Shota Rustaveli National Science Foundation grant FR17-391.
\end{acknowledgments}

\section*{Data Availability}
The data that support the findings of this study are available from the corresponding author
upon reasonable request..

%\nocite{*}
\bibliography{aipsamp}% Produces the bibliography via BibTeX.

\end{document}